# THE ORIGIN OF GALAXIES

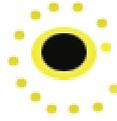

## Mass Particles, Time, Gravitation and Black Holes in Perspective of the CER Wave Packet Concept of Mass and Charge.
Author: Erik Haeffner

### INTRODUCTION

The CER concept which describes the physical origin of mass and charge has shown that mass particles are of electromagnetic origin and consist of condensed circular polarized electromagnetic radiation, EMR. It has also been shown that this theory predicts the existence of the most fundamental particles in the Standard Model, leptons and quarks of the first generation. Heavier mass particles such as mesons and baryons are found to be composed of superimposed CER wave packets. Using this knowledge it is possible to draw conclusions regarding the concept of time, the nature of gravitation and the formation of galaxies

### Mass particles as Superpositions of EMR waves

The CER concept presented in [1,2] is essentially a theory of the physical structure of mass particles and how they are produced. In interaction with certain crystals a linear polarized EMR in the light frequency band will be converted into two new, one lefthanded and one right handed, circularly polarized EMR:s. This phenomenon has been called "parametric down conversion". The physical phenomenon known as pair production is described as the process when one linear polarized EMR is converted into two oppositely charged mass particles, one electron, e-, and one positron, e+. The sum of the energies of the new created mass particles, $E = 2m_e c^2$, is the same as the energy $E=h\nu$ of the parent EMR.

The physical conclusion of these two experimentally established phenomena is that at a certain frequency (energy) of a circularly polarized EMR, the electromagnetic waves are condensed, figuratively compressed like a spiral spring, into a CER wave packet which have all the properties of a mass particle. By particle collisions at high kinetic energies in accelerators or storage rings very high energy EMR:s are created which by pair production then convert into heavy mass particles such as protons and antiprotons or positive and negative pions.

As demonstrated in [1,2] circularly or elliptically polarized CER wave packets may also be produced by the superposition of two plane polarized EMR:s at a quantum phase difference. It is found that the existence of the first generation of leptons and quarks coincides in symmetry and charge with CER particles created by the superposition of pairs of EMR :s which have a phase difference of n $\pi$ /6, where n is 0,1,2,3 and 6,7, 8,9.

If we go further with the superposition process, we can mention the mesons called pions: $\pi^+$, $\pi^-$, $\pi^0$ each as a superposition of two quarks, which each in turn is composed of two EMR:s. We find the following superpositions of CER wave packets:

$u^{+2/3} + \bar{d}^{+1/3} = \pi^+$, up quark superposed on a down antiquark,

$\bar{u}^{-2/3} + d^{-1/3} = \pi^-$, up antiquark superposed on a down quark,
$u^{+2/3} + \bar{u}^{-2/3} = \pi^0$, up quark superposed on an up antiquark.

The $\pi^0$ is a very shortlived superposition ($0.84 \cdot 10^{-16}$ s) and disintegrates into two plane polarized EMR:s. The result might be called an annihilation reaction.
According to the CER theory the proton and the neutron can each be explained as a superposition of three quark wave packets (**uud**) and (**ddu**) respectively, in which each quark is a superposition of two EMR:s as described in (1,2). The antiproton is in the demonstrated symmetry composed of two anti up quarks and one anti down quark ($\bar{u}\bar{u}\bar{d}$).

**The fundamental difference between the proton and the antiproton appears thus to be the phase difference between the initial superposed EMR:s**.
It has been shown at CERN that when protons and antiprotons collide at very high energies the mentioned different quark wave packets will react in the following ways:

$u^{+2/3} + \bar{d}^{+1/3} \longrightarrow W^{+1} \longrightarrow e^{+1} + \nu_e$
$d^{-1/3} + \bar{u}^{-2/3} \longrightarrow W^{-1} \longrightarrow e^{-1} + \bar{\nu}_e$
$\bar{u}^{-2/3} + u^{+2/3} \longrightarrow Z^0 \longrightarrow e^{-1} + e^{+1}$
$d^{-1/3} + \bar{d}^{+1/3} \longrightarrow Z^0 \longrightarrow e^{-1} + e^{+1}$

Using the CER concept these reactions can be interpreted as follows. The **w⁺¹** is an intermediate circularly polarized wave packet which disintegrates into a positron and an electron neutrino $\nu_e$ which is a superposition of two plane polarized EMR:s with no phase difference or a phase difference of $6\pi/6$. The $W^{-1}$ is also an intermediate CER wave packet which disintegrates into an electron and one anti electron neutrino $\bar{\nu}_e$, formed by two plane polarized EMR:s with no phase difference.

The $Z^0$ intermediate particle should be interpreted as an annihilation EMR which immediately converts into two mass particles $e^+$ and $e^-$ according to the CER concept and its fundamental pair production phenomenon.
**The building of mass particles may thus be illustrated and explained by the superposition phenomenon of initial EMR:s.**

## About the Concept of Time

Both classical and relativistic physics assume a time concept which is felt or experienced by humans as an infinitesimal amount or instant of time going from past through present to the future. According to classical physics the passage of time is uniform and flows at the same rate throughout the Universe.
As the classical concept of time, so far, has not got any formal physical definition, it is difficult to appreciate if other proposals of time concepts are of better value for the science of physics. Consequently there has been a great amount of speculations and analytical thinking on and about time in physical and philosophical literature.
I shall here refer to only a few of all these authors as the present approach is founded on a new theory, discussed in [1,2] regarding the material content of the universe. As an introduction and state of the art I should, however, like to mention a very informative book by Paul Davies [3]

Donald F Weitzel [4] and others have pointed out that the concept of time as it is used in physics hardly is a "basic feature". We have no knowledge of time itself in classical physics exept by the means of motion. Clocks can, however, be designed to show motion and used in the same way as time clocks, i.e. to determine and use a standard of motion. We will then get a natural and understandable combination of space and classical time which might be interpreted as a "basic entity". On these grounds we can call and characterize the time we measure with clocks in ordinary life and in physics as "**motion time**". The motion time is perceived as occuring in a three dimensional space

The General Relativity Theory found by Einstein is a mathematical model to explain the concept of gravity and its interaction with both mass particles and EMR (photons) which according to the Standard Model is assumed to have no mass. (In the QM theory the photon is a wave as well as a point particle, that is a point particle without a mass or, if you will, without a point. To put it drastically: there is no point in talking about a point particle without a point, so I have in the name of clarity refrained from talking about a "photon" in any other meaning than the energy quantum **E = h$\nu$**)

In order to find equations that would correspond to experimental evidence and conform with observations of natural phenomena it was necessary to create a new concept called "spacetime", and assume that gravity, in fact, is a warping of the four-dimensional spacetime. This is of course a very elegant mathematical solution to the phenomenon of acting at a distance.

As physical evidence for the time warping, a few physical phenomena have been metioned, among them the deflection of light, EMR, when passing near the sun, that is a high gravity field, and establishing that clocks are going faster at higher altitudes. In an extreme gravitation, such as near a black hole, the warping of spacetime is also so extreme that EMR:s cannot leave the "event horizon" of the black hole.
With the introduction of the CER concept, however, the following properties of EMR seem clear:
- a photon (EMR) is not a mass particle but it is the containment of an energy quantum **E = h$\nu$**; (Planck - Einstein)
- by circular polarization and condensation into a wave packet the photon is converted into a CER mass particle, **E= h$\nu$ =m c$^2$**,
- the Plancks constant **h** can be interpreted as the energy content of one wave unit **E$_u$**

$$E_u = h = m_u c^2, \quad m_u = h/c^2$$

which as a mass unit, ($m_u$) follows the gravitation laws and is consequently deflected in a gravity field. With this knowledge, time as a fourth dimension in a spacetime universe warped by gravity does not seem to be a necessary hypothesis to explain physical facts and phenomena such as the deflection of EMR by the sun. We can therefore call this time concept as a **"mathematical time"**.
For a discussion of the deflection angle if we use the equivalent mass of the photon m = E/c$^2$ we can refer to L.A. Popedonostev [6].

As expressed by Feynman [5] there are two laws of physics which apply to cosmos in a general way.

**First law: The energy of the Universe is (always) constant.**

**Second law: The entropy of the Universe is (always) increasing.**

The word "always" in these statements is put inside parentheses because it indicates the existence of a universal steady time concept, which so far is just a "feeling" or assumption which has not been proved to exist. For the following discussion the word "always" in the two laws above will therefore not be taken into account.
It was shown in [1,2] that the content of the universe is of electromagnetic origin either as **E = h$\nu$ or E = h$\nu$ = mc$^2$**

(We use this energy formula according to Einstein in comparing the energy of h$\nu$ with mc$^2$. In the production reaction of mass particles from EMR we note, however, that h$\nu$ =

2 mc² where hν is the energy of an EMR and **m** is the mass of a CER wave packet as demonstrated by the predictions of charge and symmetry for the first generation leptons and quarks.[1,2] )
This discovery allows us to express an energy formula for the content of the whole Universe

$$\Sigma \ h\nu_i + \Sigma \ m_j c^2 = Ku$$

**Ku = the constant energy of the universe according to the first law.**

As the total energy is constant there is no time differential involved in the first law. The second law, however, means an irreversional uni - directional change, and this fact has inspired several scientists to use the entropy concept for the measure of time.
The change of entropy in a system is generally defined as dS = dQ/T where dQ is an amount of heat (energy) and T is the absolute temperature which, in fact, is a convention from the kinetic gas theory

$$m v^2 / 2 = k T$$

where
k = Boltzmans constant.
m = particle mass and
v = particle velocity

From this definition it might be said that the entropy change is the energy amount it takes per temperature degree Kelvin to mix the particles in a system from the lower to the higher entropy level. This irreversible change might be measured in isolated thermodynamic systems but for the whole universe how can we measure the overall change of the entropy level? This seems, however, to be as near as we can come a "cosmological time" scale, starting when the entropy of the initial system starts to increase. What was the initial system? The Big Bang theory has been a favourite idea for some time. A black hole might be another starting object, an especially inviting speculation since Stephen Hawking [ 7 ] calculated a probability that black holes evaporate high energy gamma radiation.
**Summarizing this section we do not find any definable physical entity of time, but we find two concepts that have properties that conform with the human feeling of an existing flow of time, one is the "motion time" as described above, and the other is the concept of entropy increase which can be appreciated as a "cosmological time scale".**

## The Origin of Galaxies

We know from [1,2] that mass particles are produced from circularly polarized and condensed electromagnetic radiation called CER wave packets. The conclusion is that all materia in the universe is of electromagnetic origin. Also black holes must then contain highly compressed electromagnetic radiation which creates an extreem gravity field strength. Stephen Hawking using QM mathematics, has proposed that some high energy EMR will, in spite of this gravity field, in consequence of the Heisenberg uncertainty principles escape from the black hole and pass through the event horizon. The Heisenberg uncertainty pairs, which are the foundation of QM, are also deduced in the CER/Haeffner theory as shown in [1]. The difference is of course that QM does not refer to EMR wave packets but to physically unidentidfied mathematical waves or to matter waves, whatever that means.

Through known down-conversion reactions the EMR escaping from a black hole will be converted to mass particles and lower energy EMR. A cloud of particles, a galaxy, will be formed surrounding the initial black hole.

The conclusion is that each galaxy is created by a black hole and the origin or, if we use the "cosmic time" concept, the beginning of the universe is one big black hole which, as proposed by Hawking, might have been divided and the pieces spread out to form many smaller holes.

The spiral structure of galaxies is a beautiful illustration how the CER concept functions in the Universe. High energy electromagnetic radiation emitted by a black hole is converted into mass particles by two physical procedures.

**The first** is realized when one plane polarized EMR is converted into two circular polarized EMR:s which eventually, if the frequency is high enough, are condensed (compressed) into electromagnetic wave packets. These wave packets,CER, are in all respects identical with what we call mass particles in pair production processes.

**The second** way is the superposition of two or several EMR:s with a phase difference which produce CER wave packets. As an example, the superposition of two EMR:s at a phase difference of $\pi/6$ will produce successively the first generation of leptons and quarks in the Standard Model.[ 1 ]

These two physical processes take place in the spiral arms of galaxies where new mass particles consequently are created. If a black hole is rotating anticlockwise and the EMR:s are emitted radially, these will be bent in the high gravity field to a spiral structure opened in the clockwise direction. That is also what we see in countless cases in the Universe.

During the last few years a very high activity in astrophysics has been carried out by means of telescopes both on earth and in space in order to find new information leading to an understanding of the origin of galaxies. An excellent compilation of this new knowledge with comments on surveys regarding different kinds of electromagnetic background radiations has ben made by Günther Hasinger and Roberto Gilli. [ 8 ] A few of these recently discovered realities seem difficult to understand on the basis of the Big Bang theory or on generally accepted physics. It is therefore especially interesting to discuss these dubious points in the light of the CER solution, just mentioned, to the problem of galaxy formation. Astronomers have found a galaxy, NGC 6240, which seems so undisturbed that it can be used as a model for other galaxies´ formation and life. The overall spectrum of NGC 6240 has the same shape as that of the cosmic background electromagnetic radiation in five observed frequency bands. Moreover this galaxy has a high production of stars as well as a huge black hole. These two phenomena seem to have a very high correlation of appearence in time and space, which so far has been a mystery for astronomers. In the light of the CER theory the close relation between a black hole and the surrounding galaxy is self-evident as mass particles and materia are formed by gamma radiation emitted from black holes.

During the last decades and paticularly during the last two years astronomers have made great efforts to find and study background electromagnetic radiation in different wavelengths. The earliest discovered background radiation, is the CMB, cosmic microwave background radiation, at a photon energy of $10^{-3}$ eV. Recent observations have indicated that the CMB contains circularly polarized radiation as predicted by the CER theory.

High energy X -radiations at $10^5$ eV, CXB, were first found to be uniformly spread over the sky by detectors on satellites, such as the Chandra X-ray Observatory. This indicates an origin outside our solar system as well as outside our galaxy. On the other hand, surveys made at a lower energy, up to 10 keV, by the ROSAT X-ray satellite claim that about 80% of the X-ray sources are identified as active galactive nuclei of various kinds such as luminous quasars or so-called Seyfert - 1 galaxies containing black holes.

Background radiation is now also found as CGammaB, at very high energies, $10^9$ eV. At lower energies: COB, Cosmic Optical Background, at ca 2 eV, and CIB, Cosmic Infrared Background at ca 1-10 eV. have also been detected.

It is assumed that only high energy gamma radiation can leave a black hole because of the high gravity field. That means that lower energy background radiations are formed by processes outside the black hole through energy down conversion reactions. Well-known and experimentally verified paths are pair productions such as proton and antiproton ; electron and positron ; electron and electron neutrino; Compton radiations, etc.

From the home of the solar system in the Universe it has not been an easy task to find all the background cosmic radiation mentioned here, among lots of other EMR of a secondary origin, e g from fusion reactions in stars or X radiations from intergalactic gas and dust.

There are three main conclusions from later years astrophysical surveys and observations which support the prediction in this article that black holes are the origin of galaxies.

1. It is observed that black holes and galaxies appear in the same place at the same cosmic time which indicates that they are somehow related. Large black holes seems to be a prerequisite for a high star production in the galaxy arms.
2. The galaxy arms follow the trajectories of high energy gamma radiation from AGN:s (Active Galactic Nuclei, black hole.) The structure of the galaxy is an Archimedes spiral, the geometry of which depends on the rotation of the black hole and its gravity field strength.

The recent discovery of the cosmic background radiation in different electromagnetic wavelength bands conforms with the theory that the initial gamma radiation from a black hole is transformed to lower energy radiation by mass particle production and other down conversion reactions, until the CMB.

## Conclusive Evidence for the Physical Validity of the CER Concept

Established physical facts and experimental results have been mentioned in earlier articles [1,2] as evidence for the new concept of condensed electromagnetic radiation, CER. In order to get an overview of the impact of this new theory for the understanding of well-known physical phenomena, as well as for the content of the Universe, a list of 9 subjects of importance for the science of physics today is discussed in the following section. As evidence, each point contains a comparison between a prediction or statement which can be made from the grounds of the CER theory and established physical facts on the other side.

**Prediction 1.** From a phenomenon called "parametric down conversion", that is when a linear polarized EMR passing a crystal structure is converted into two new circular polarized EMR:s - one with lefthanded polarization and the other in a symmetrical righthanded polarization, it is predicted that at a certain higher frequency (energy) of the initial EMR the secondary circularly polarized EMR:s are condensed (compressed) into CER wave packets, that is positively resp. negatively charged mass particles.
**Physical facts:** The pair production phenomenon is a well-known and experimentally established reaction which exactly conforms with the mentioned prediction in the production of electrons and positrons from a high energy EMR.

---

**Prediction 2**. The leptons and quarks of the first generation are each formed by the superposition of two linear polarized EMR:s with a quantum phase difference of $\pi/6$ radians.

**Physical facts**: The experimentally confirmed existence of the first generation leptons and quarks conform with the predicted CER particles in symmetry and charge.

---

**Prediction 3.** The charge on mass particles depends on the quantified polarization degree and the spin direction of their CER wave packets.
**Physical facts:** The superimposed CER particles coincide in charge with the generally accepted existence of the first generation leptons and quarks.

---

**Prediction 4.** A CER wave packet which conforms with the electron in symmetry and charge should naturally have a characteristic wavelength when used in Compton scattering experiments.
**Physical facts:** The electron mass particle shows in Compton scattering experiments a characteristic wavelength. The same wavelength appears in de Broglie diffraction experiments.

---

**Prediction 5.** Two linear EMR:s with a phase difference of 2 $\pi$/6 form an up quark antiparticle $\bar{u}$ with the charge - 2/3. Additional two EMR:s with a phase difference of **7 $\pi$/6** form a down quark, antiparticle, $\bar{d}$. As CER wave packets these can superimpose to form a mass particle $u\bar{d}$ called a $\pi^+$ meson. In the same way a proton may be formed out of originally 6 linear EMR:s, primarily with phase differences in pairs to form quarks. With other combination of phase differences other mass particles, antiprotons and neutrons are formed.
**Physical facts:** By collision experiments $\pi$ mesons have been found to contain one u quark and one $\bar{d}$ quark. Protons have been found to contain three quarks, (uud) as the antiprotons ($\bar{u}\bar{u}\bar{d}$ as well as the neutron (ddu).

---

**Prediction 6.** The CER wave packet concept can be seen as a physical alternative to the quantum mechanical wave equations, when we predict that the structure of matter is of electromagnetic origin. On the other hand we leave the QM point particle concept of mass as a mathematical model without a physical reality.
**Physical facts:** The Heisenberg uncertainty principles which are the foundation of the QM theory can also be deduced as shown in [1] by using the CER concept of electromagnetic wave packets. **The duality between a mass particle and a wave packet is in this concept obvious, natural and physically real**.

---

**Prediction 7.** We can consider the photon h$\nu$ as an energy quantum spread out along the trajectory of the electromagnetic radiation. The Planck constant **h** can be interpreted as the energy of one wave unit which according to Einstein has an equivalent mass unit $m_u = h / c^2$ Using this mass unit we can predict that even in a 3D universe an EMR is deflected by a gravity force.
**Physical facts:** It has been demonstrated experimentally that EMR is deflected by gravity. This phenomenon has been explained and described by the Einstein General Relativity 4 dimensional mathematical model. By the introduction of the CER theory which implies a mass equivalence of EMR there is no need to assume a warped spacetime model to explain the deflection of light and other electromagnetic radiation in gravity fields.

---

**Prediction 8.** Galaxies are created by black holes radiating EMR. which, above a certain frequency (energy), according to the CER concept produce new mass particles by superposition and pair production reactions.
**Physical facts:** There are no other mass creation reactions known in experimental physics than the CER condensation of circularly or elliptically polarized EMR:s into mass particles and initially by superposition of linear polarized EMR:s with a phase difference. Stephen Hawking has also, from QM mathematical deductions, proposed that black holes are emitting electromagnetic radiations that may produce electrons and positrons by pair production. (The physical prerequites for the Hawking propositions are not supported by current theories. The General Theory of Relativity does not allow EMR:s to escape from black holes do to the warping of spacetime. The QM theory does not include the possibility that "matter waves" or "mathematical " wave packets are of electromagnetic origin.)

---

**Prediction 9.** If the amount of energy in the Universe is assumed to be a constant, given once and forever, there is no change of energy against a conceivable universal time concept which thus has no meaning. The human mind experiences a flow of time on which historic events, the instants of now and the future may be fixed. This human time consciousness has found a physical interpretation in the concept of motion, clocks and similar devices, which serves us well, but cannot be defined for cosmic reference.The nearest we can come a cosmological time concept is the irreversible increase in entropy. It is predicted that only two time concepts have a physical identity, one is the **motion time** which is of value and use for human activities and scientific calculations, the other is the **cosmological time**, which is defined as the total increase in the entropy of the universe.
**Physical facts.** An overall irreversible increase in entropy is generally accepted as a reality. It has been suggested by Hawking et. alt.[7] that the surface inside the event horizon of a black hole has the dimensions of entropy. The starting point of cosmological time would then be the division of one first big hole into many smaller black holes, and in this way increasing the entropy and the flow of cosmological time.

**References**

[1] E.A. Haeffner, "The Physical Origin of Mass and Charge", http://www.algonet.se/~haeffner, Feb. 1, 1997 (Officially registered June 2, 1995)
[2] E.A. Haeffner, "A Physical Origin for Mass and Charge" 112-115 Galilean Electrodynamics, Nov/Dec 2001.
[3] Paul Davies, "About Time", Penguin Books Ltd, London 1995
[4] Donal F. Weitzel, "Time: The Shadow Dimension", Galilean Electrodynamics, Special Issue, 1999.
[5] R.P. Feynman, R.B. Leighton, Matthew Sands. "The Feynman Lectures in Physics" Addison - Wesley Publishing Co INC, Reading, Massachusetts, 1966.
[6] L.A. Pobedonostev "On the Mass of the Photon", Galilean Electrodynamics, Vol. 9,Number 5, pp 98-99.
[7] Stephen Hawking, "Black Holes and Baby Universes and Other Essays", Bantam Press, London 1993, pp 90 - 103 .
[8 ] Günther Hasinger, Roberto Gilli "The Co Cosmic Reality Check" Scientific American Vol 286, No 3, pp 46-53.